\documentclass[aps,prd,superscriptaddress, 11pt]{revtex4}

\usepackage{commath}
\usepackage{graphicx}
\usepackage{epstopdf}
\usepackage{amsmath}
\usepackage{amsfonts}
\usepackage{amssymb}
\usepackage{latexsym}
\usepackage{hyperref}
\usepackage[english]{babel}
\usepackage[utf8]{inputenc}
\usepackage[colorinlistoftodos]{todonotes}
\usepackage{xcolor}
\usepackage{slashed}
\usepackage{bm}  % To make greek letters in bold face {\bm \Phi}
\usepackage[normalem]{ulem} %tabela
\usepackage{amsmath,mathtools}
\usepackage{caption}
\usepackage{subcaption}
\usepackage{float}
\useunder{\uline}{\ul}{}
\usepackage{hyperref}
\usepackage{layout}
\begin{document}

\author{Jos\'e J. Ram\'on Mar\'i\footnote{Corresponding author.}}
\email{jjramon@cbpf.br}
 \affiliation{Instituto Militar de Engenharia (IME),\\
Pra\c ca Gen. Tib\'urcio 80, Urca, Rio de Janeiro, RJ, Brazil, CEP 22291-270 } 

\author{J. A. Helay\"el-Neto}
\email{helayel@cbpf.br}
\affiliation{Centro Brasileiro de Pesquisas F\'{i}sicas (CBPF),\\ Rua Dr. Xavier Sigaud 150, Urca, Rio de Janeiro, RJ, Brazil, CEP 22290-180}

\author{Y.M.P. Gomes} 
\email{ymuller@cbpf.br}
\affiliation{Centro Brasileiro de Pesquisas F\'{i}sicas (CBPF),\\ Rua Dr. Xavier Sigaud 150, Urca, Rio de Janeiro, RJ, Brazil, CEP 22290-180}

\newcommand{\be}{\begin{equation}}
\newcommand{\ee}{\end{equation}}
\newcommand{\bea}{\begin{eqnarray}}
\newcommand{\eea}{\end{eqnarray}}
\newcommand{\bean}{\begin{eqnarray*}}
\newcommand{\eean}{\end{eqnarray*}}
\newcommand{\I}{1\!\mbox{I}}
\newcommand{\Q}{\mathbb{Q}}
\newcommand{\R}{\mathbb{R}}
\newcommand{\C}{\mathbb{C}}
\newcommand{\zz}{\mathbb{Z}}
\newcommand{\go}{\omega}
\newcommand{\ga}{\alpha}
\newcommand{\gb}{\beta}
\newcommand{\pp}{\mathbb{P}}
\newcommand{\ff}{\mathbb{F}}
\newcommand{\af}{\mathbb{A}}
\newcommand{\cO}{\mathcal{O}}
\newcommand{\cM}{\mathcal{M}}
\newcommand{\cH}{\mathcal{H}}
\newcommand{\nn}{\tilde{\mbox{n}}}
\newenvironment{defn}[1][Definition]{\begin{trivlist}
       \item[\hskip \labelsep {\bfseries #1}]}{\end{trivlist}}

\newtheorem{theorem}{Theorem}[section]
\newtheorem{prop}[theorem]{Proposition}
\newtheorem{corollary}[theorem]{Corollary}
\newtheorem{cor}[theorem]{Corollary}
\newtheorem{lemma}[theorem]{Lemma}
\newtheorem{obs}[theorem]{Remark}
\newtheorem{exc}[theorem]{}
\newtheorem{example}[theorem]{Example}

\newenvironment{remark}[1][Remark]{\begin{trivlist}
       \item[\hskip \labelsep {\bfseries #1}]}{\end{trivlist}}

\newenvironment{caja}[1]{\begin{trivlist}
       \item[\hskip \labelsep {\bfseries #1}]}{\end{trivlist}}

\title{\bf F-term spontaneous breaking of 3D-SUSY: an algebro-geometric treatment}

%\begin{document} 

\begin{abstract}
    We settle a result on the generic exactness of simple SUSY in 3D, and provide a mechanism of F-term spontaneous breaking, with a different set of tools from those adopted by O'Raifeartaigh in his seminal work on the F-term spontaneous breaking of 4D SUSY. 
In our study, we adopt techniques of projective algebraic geometry so as to deal successfully with the cubic hypersurfaces involved.
\end{abstract}
%\maketitle

\maketitle
\flushbottom

%%%%%%%%%%%%%%%%%%%%%%%%%%%%%%%%%%%%%%%%%%%%%%%%%%%%%%%%%%%%%%%%%%%%%%%%%%%%%%%%%%%%%%%%%%%%%%%%%%%%%%%%%%%%%%%%%%%%%%%%%%%%%%%%%%%%%%%%%%%%%%%%%%%%%%%%%%%%%%%%%%%%%%%%%%%%%%%%%%%%%%%%%%%%%%%%%%%%%%%%%%%%%%%%%%%%%%%%%%%%%%%%%%%%%%%%%%%%%%%%%%%%%%%%%%%%%%%%%%%%%%%%%%%%%%%%%%%%%%%%%%%%%%%%%%%%%%%%%%%%%%%%%%%%%%%%%%%%%%%%%%%%%%%%%%%%%%%%%%%%%%%%%%%%%%%%%%%%%%%%%%%%%%%%%%%%%%%%%%%%%%%%%%%%%%%%%%%%%%%%%%%%%%%%%%%%%%%%%%%%%%%%%%%%%%%%%%%%%%
\section{Introduction}

     Supersymmetry (SUSY) appears in relativistic field theories (both classical and quantum) as a space-time symmetry with the remarkable consequence of  placing bosons and fermions in the same (linear) representation space, called supermultiplet. This is why it is usually referred to as a fermion-boson symmetry.  Besides being an elegant solution to the naturalness and gauge-hierarchy problems of the Standard Model of Particle Physics, SUSY also implies that the running coupling constants of the electroweak and strong interactions can be unified at a high energy scale, $10^{16}$ GeV, providing then a viable path towards unification.
     Supersymmetric partners of the particles of the Standard Model have however not yet been observed at available accelerator energies (LHC, in Run II, has been exploiting physics up to 13 TeV centre-of-mass energy). Therefore, to describe some new physics beyond the Standard Model, SUSY needs to be broken by some particular mechanism (be it explicit by soft terms, spontaneous or dynamical). Even if the breaking should occur at very high energies, it can be communicated to the low-energy sector of the spectrum \cite{MSSM}.
     In four space-time dimensions, as is well-known, SUSY exhibits a complex structure: the fundamental representation of simple four-dimensional (4D) SUSY is realised in terms of complex superfields and their associated complex component fields. That is a crucial point in connection to a particular mechanism of SUSY spontaneous breaking, namely, F-term SUSY breaking, thoroughly studied by O'Raifeartaigh in his classical 1975 papers \cite{ORafer}. In \cite{ORafer}, the author turns the analysis of F-term spontaneous simple SUSY breaking into the study of a system of N quadratic equations in N complex unkowns (these N unkowns being space-time constant configurations of N complex scalar fields accommodated in N matter superfields), and shows that spontaneous breaking is possible only if there are at least three superfields that display these scalar fields. The results of \cite{ORafer} gave rise to the so-called O'Raifeartaigh models. (Just to avoid any possible misunderstanding: N stands for the number of fields present in the particular model under consideration. Here we shall be dealing only with simple SUSY, and so N never refers to the number of extended supersymmetries.)
  
		SUSY has been carefully explored in diverse space-time dimensions (more specifically, from two to eleven). In particular, SUSY in three space-time dimensions received a remarkable boost in relation to Chern-Simons and planar field-theoretic models \cite{CSTFT3D}. More recently, renewed interest in 3D SUSY has arisen in connection to topological materials in lower space dimensions,  for instance (1+2)-dimensional topological superconductors \cite{topsemicond}, where SUSY appears as an emergent symmetry of the action which describes the dynamics of the excitations.  So, motivated by the relevance that SUSY is nowadays acquiring as an emergent symmetry in low-dimensional Condensed Matter systems such as graphene, topological insulators and topological superconductors, we re-examine the issue of F-term SUSY spontaneous breaking in three space-time dimensions.

  Contrary to the 4D case, (1+2)-D minimal SUSY has a real structure: simple SUSY in three space-time dimensions is realised in terms of real superfields and their corresponding real component fields. Renormalisation requirements in 3D allow the matter superfields to have up to quartic interactions, whereas in 4D, the coupling amongst matter superfields is at most cubic. In terms of component fields 3D-SUSY, in connection with renormalisability, allows a potential which is of 6-th order in the scalars; on the other hand, 4D-SUSY allows scalars up to fourth power. This drastically changes the structure underneath spontaneous SUSY breaking in comparison to its four-dimensional counterpart: instead of a system of N quadratic equations in N complex unkowns, the conditions for 3D F-term spontaneous SUSY breaking hinge on the existence of real solutions of a system of N cubic equations in N real variables, which renders reference to the seminal results in \cite{ORafer} no longer suitable, and calls for different techniques.  We carry out the analysis of the corresponding system of cubic equations derived from F-term breaking using  projective algebraic geometry.  Today, algebraic geometry is no stranger to theoretical physics; see the work of Candelas et al \cite{Candelas} for a recent example; for a sample of applications to biology, see \cite{Casanellas}. Here we use classical tools from this field to solve a physical question.

  Explicitly, (1+2)-D minimal SUSY with N superfields (pure matter fields) is described by a superpotential, namely, an arbitrary real polynomial $V(\phi_1, \cdots, \phi_N)$ of degree at most 4 in the superfields $\phi_i$ (which are real scalars).  We analyse the system of algebraic equations (of degree at most 3) given by $\frac{\partial V}{\partial \phi_i}=0$ (traditionally called $F_i$ or `F-term'), i.e. we study the existence of critical points of this potential. Should any exist, our SUSY is exact, otherwise it breaks spontaneously at the F-term.  The case $N=1$ amounts to the basic result that any polynomial of odd degree has at least a real zero: one of the proofs consists of applying the Fundamental Theorem of Algebra, and realising that not all zeros come in pairs $\alpha, \overline{\alpha}$.  This result no longer holds if there is degeneration to a lower degree, e.g. when the 1-parameter family of cubic equations $tx^3+x^2+1=0$ reduces when $t=0$ to a lesser degree, so our exactness theorem is bound to hold only generically.  Since we deal with an arbitrary number of variables, our arguments cannot be this simple, and that is where complex projective algebraic geometry comes in, for it is the natural context with solid counting tools.
	
   It should be emphasized that we do not attempt a phenomenological discussion here, but stick to the mathematical problem underlying F-term spontaneous breaking of SUSY, although we keep in mind applications to emergent SUSY in planar condensed matter phenomena, which shall be addressed elsewhere. We would like to highlight that we are presenting here a general demonstration in the context of  \text(1+2)-D minimal $\mathcal{N}=1$ SUSY breaking by means of an F-term. Here, in the 3-dimensional case, we are focussing on the F-term. In 3-D, and there is no analogue of a D-term. The 3D counterpart of the D-term is spinorial, and if it should break SUSY then Lorentz Symmetry would be broken as well.  Some particular examples will be presented after our generic result is set up.

     The outline of our paper is as follows: after presenting some background on algebraic geometry in Section \ref{intro}, we provide a result on generic exactness in Section \ref{exatagen} ; next, in Section \ref{explicitbreaking}, we introduce an explicit family of potentials with F-term SUSY breaking, and provide a different example, this time numerical, of an even potential with F-term breaking in Section \ref{evenpot}. Finally, in Section \ref{SUSYquebra32} we derive a small hypersurface of potentials satisfying F-term SUSY breaking in the 3-superfield case (with a $34$-dimensional ball as their parameter space) and its N-superfield analogue, building upon Section \ref{evenpot}, then cast our Concluding Comments in Section \ref{conclusion}.

\begin{caja}{Notation and conventions:}    We denote by $E=E_3$ the linear space of quartic scalar potentials $\Phi=\Phi(x,y,z)\in E$ on $\R^3$ , and by $(F,G,H)$  the gradient of $\Phi$ in the N=3 case ($E_N$ will be its $N$-superfield analogue, namely the vector space of quartic polynomials in N variables).  Note that $E$ has dimension $1+3+6+10+15=35=\binom{4+3}{3}$.  We denote by $\R_d[x_1, \ldots , x_n]$ the space of polynomials in the variables $x_i$ of degree at most $d$, and denote by $\R[x_1, \ldots , x_n]_d$ the space of homogeneous polynomials in $x_1, \ldots , x_n$. Given a polynomial $F$ in $x_i$, its homogenized form is denoted by $\tilde{F}$, i.e. $\tilde{F}(X_0,X_1,X_2,X_3)=X_0^{deg(F)} F(X_1/X_0,X_2/X_0, X_3/X_0)$ (we understand that $x_i=X_i/X_0$).

  Finally, while it is customary to assume that the superpotentials $V$ are such that $V(0)=0$, we shall not do so for reasons of convenience in our computations.  This also means that the families of superpotentials considered will have one extra dimension due to the constant terms (or two, for those preferring to work in $\pp(E_N)$ instead of $E_N$).\end{caja}

\section{Hypersurfaces in $\pp^n_{\C}$ and their intersection zero-cycles} 

\label{intro}

    See \cite{Sha} (especially \cite[I Ch. 4]{Sha}) \cite{GH} for details beyond this summary.  The purpose of this section is to provide the toolbox to be used in the proof of our main results on real solutions. Basically, if we have a finite set $S$ of complex points in $\pp^r_{\C}$ which is invariant under complex conjugation and its cardinal is odd, then $S$ must contain a real point -- our arguments shall be similar, albeit counting multiplicities.  The most elementary case is that of the intersection of two plane curves, beautifully treated in \cite[3.3]{Fulton} and in \cite[I, Ch. 1]{Sha} (see also \cite[Ch.1]{Bix} for an elementary, undergraduate-oriented presentation).
		
		We now set ourselves in the complex projective $n$-space $\pp^n_{\C}=\pp^n,$ with projective homogeneous coordinates $X_0, \ldots , X_n.$  The affine coordinates are by default $x_i=\frac{X_i}{X_0}$, and a \textbf{form} of degree $d$ is a homogeneous polynomial of degree $d$. We start with a simple situation in the complex plane: clearly, a complex projective line $\ell\subset\pp^2$ given by $L=0$ and a curve $F=0$ (where $F(X_0,X_1,X_2)$ is a form of degree $d$ in the $X_i$'s) intersect in $d$ complex points, counting multiplicities, unless $L$ divides $F$.  Likewise, we see that two conics with no common factors $Q_1,Q_2$ intersect in $2\times 2 = 4$ points counted with multiplicities, which we write in intersection product notation as follows:  $\{Q_1=0\}\bullet\{Q_2=0\}=\{Q_1=0\}\bullet\{Q_3=0\},$ where $Q_3=Q_2+\lambda Q_1$, where $\lambda$ is a solution of $det(Q_2+x Q_1)=0$, so $Q_3$ decomposes into linear factors $Q_3=L'L$, i.e. (denoting $\ell=\{L=0\}, \ell'=\{L'=0\}$) 
$$\{Q_3=0\}=\ell+\ell'.$$  Thus, 
$$\{Q_1=0\}\bullet\{Q_2=0\}=\{Q_1=0\}\bullet (\ell+\ell')=\{Q_1=0\}\bullet\ell+\{Q_1=0\}\bullet\ell'=2+2=4.$$.
  B\'ezout's Theorem in $\pp^2$ generalises this to the following result: if $F,G$ are forms of respective degrees $d,e$ in the variables $X_0, X_1, X_2$, and have no common factors, their total intersection number in $\pp^2$ (i.e. counting multiplicities) is $(F=0)\bullet (G=0)= F\bullet G=de$ (we abuse notation identifying the locus with its equation).
  
  In the case of $\pp^n,$ one has $D_1, D_2, \ldots, D_n$ hypersurfaces, where $D_i$ is the zero locus (scheme) of zeros of a form $F_i$ in $X_0, \ldots, X_n$ of degree $d_i$.  Assume that they properly intersect, i.e., that their intersection $F_1=0, \ldots , F_n=0$ is a finite subset of $\pp^n_{\C}$.  One may define an intersection $0$-cycle $$D_1\bullet \cdots\bullet D_n= \sum_{P\in\pp^n} \mu_P P,$$
 where $\mu_P$ is the intersection multiplicity of $F_1, \ldots, F_n$ at the point $P\in\pp^n$, and all but a finite number of $\mu_P$ are zero.  If ${\mathcal O}_P$ is the ring of regular functions at $P$ and $f_i=0$ are local equations defined by $D_i$ (e.g. taking $f_i(x)=F_i(1,x_1, \ldots, x_n)$ if $P$ is in $X_0\neq 0$) one has:
 $$\mu_P=\mu_P(F_1, \ldots, F_n)=\dim_{\C}\frac{{\mathcal O}_P}{(f_1, \ldots , f_n)}.$$
\begin{obs} Serre's GAGA \cite{GAGA} shows that we obtain the same result if ${\mathcal O}_P$ is taken to be the ring of germs of holomorphic functions of $\pp^n$ at $P$, as both rings have the same completion.
\end{obs}

  The following Lemma will be used throughout.
\begin{lemma}\label{realmult} Let $F_1, \ldots F_n$ be real forms on $\pp^n_{\C}$ intersecting properly, and let $P, \overline{P}$ be a complex conjugate pair of zeros.  One has 
$$\mu_P(F_1, \ldots, F_n)=\mu_{\overline P}(F_1, \ldots, F_n).$$
\end{lemma}
{\bf Proof:} Take a suitable affine open set $X_i\neq 0$ containing $P$, and use the corresponding affine coordinates therein (which we name $x_k$).  Given a complex polynomial $f=\sum a_I x^I$ (using multi-index notation for the monomials), sending $f$ to its complex conjugate $\overline{f}=\sum \overline{a_I} x^I$ defines is an antilinear isomorphism which carries out to $\cO_P\to \cO_{\overline{P}}$, and induces a $\C$-antilinear $\R$-algebra isomorphism of the corresponding quotients 
$$\frac{\cO_P}{(f_1, \ldots , f_n)} \to\frac{\cO_{\overline{P}}}{(f_1, \ldots , f_n)}.$$
  The assertion follows. $\blacksquare$

  Let us summarise the case of $\pp^n$.  Use now the correspondence between hypersurfaces (i.e. their Weil divisors) and (isomorphism classes of) line bundles, whereby a hypersurface of degree $d$ is seen as a nonzero global section of the line bundle $\cO_{\pp^n}(d)$ \cite[I, Sec. 3.1]{Sha} \cite[Sec. 1.1]{GH}.  The intersection product as defined above yields a symmetric $n$-linear map on the Picard group $\mbox{Pic }\pp^n_{\C}$ (i.e. the group of isomorphism classes of line bundles on $\pp^n$, see \cite[I, Ch. 4]{Sha} for a full account), and since the hypersurface $D_i$ is linearly equivalent to $d_iH_i$ ($H_i$ being any hyperplane), where $d_i$ is the degree of $D_i$, their divisor classes are equal, and so is the total degree of $D_1\bullet\ldots \bullet D_n$ is (cf. \textsl{loc. cit.})
	\be\label{bezoutsth}
	\sum n_i=\deg D_1\bullet\ldots \bullet D_n=\deg (d_1\cdots d_n) H_1\bullet \ldots \bullet H_n=d_1d_2\cdots d_n \mbox{  \textbf{(B\'ezout's Theorem)}},
	\ee
	where we take $H_i$ to be generic hyperplanes, and the intersection cycle $H_1\bullet \cdots \bullet H_n$ corresponds precisely to a point with multiplicity one, since their intersection is transversal.
	
\begin{obs}	Formula (\ref{bezoutsth}) furnishes an upper bound to the cardinal of the zero set of the forms $F_i$ above, whenever finite.\end{obs}

\begin{example} An extreme example of plane cubics having one real intersection point and none in the affine plane is that of the equations $y^2=f(x)$ and $y^2=f(x)+1$ (here $x=X_1/X_0$, $y=X_2/X_0$). In this case, after homogenisation they intersect only at the flex $(0:0:1)$ with multiplicity $9$.  Indeed, let $F=Y^2Z-Z^3f(X/Z)$; we seek to compute 
$$F\bullet (F-Z^3)=F\bullet Z^3=3 F\bullet Z= 9(0:0:1)$$ 
(see \cite{Fulton}).  Thus Bezout's $9$ credits are all spent on one point, which lies at infinity.  This example may be generalised to the case of three cubic surfaces in projective $3$-space, see for instance Theorem \ref{familiapasso1}.\end{example}

  While $n$ forms on the variables $X_0, \ldots, X_n$ must have a common zero in $\pp^n_{\C}$, this is not the case if we add one more form, which generically yields an overdetermined system.  In this case, existence of such solution entails an algebraic condition (a kind of higher resultant, generalising Sylester's original) on the coefficients of the forms involved.
	
\begin{prop} \label{discr} Given the vector space of forms of degree $d$ in the indeterminates $X_0, \ldots, X_n$, there is a hypersurface ${\mathcal D}$ in $\pp(\C[X_0, \ldots, X_n]_d)=\pp^{\binom{n+d}{d}-1}$ (called \textbf{discriminant locus}) which paremetrizes all non-smooth hypersurfaces of degree $d$ in $\pp^n$.

  Analogously, given a set of $n+1$ forms $G_i$ in the variables $X_0, \ldots, X_n$, there is a multi-homogeneous polynomial $R(G_1, \cdots, G_{n+1})$ in the coefficients of the $G_i$'s, homogeneous in each argument, which vanishes if and only if the zero set of all $G_i$ is non-empty. \end{prop}
{\bf Proof:} See \cite[I.6. Exercise 10]{Sha} (see also \cite[Sect. IX.4]{Lang} and \cite[pp.9-12]{vdWaerden}).  One may in fact prove using the references cited (or the Galois theory in \cite[Ch.I]{Silverman}) that the resulting hypersurface has rational coefficients in the coefficients of $F$ (resp. the $G_i$'s). $\blacksquare$

\begin{prop}[``Liouville's Theorem''] \label{liouville} Let $Z\subset\pp^N$ be the zero set of $r$ arbitrary forms $F_1, \ldots, F_r.$ If $Z$ is not finite, then for any hyperplane $H$ the intersection $H\cap Z$ is non-empty.\end{prop}
{\bf Proof:} It suffices to apply \cite[Th.I.1.22, p.70]{Sha}. $\blacksquare$

\begin{prop}\label{zeromeasure} A hypersurface $f(x_1, \ldots, x_n)=0$ in $\R^n$ or in $\C^n$ has Lebesgue measure zero.\end{prop}
{\bf Proof:} An easy way to prove this is by using the Noether Normalisation Lemma, i.e. applying a suitable linear base change so we may assume that $f(x_1, \ldots, x_n)=x_1^d+a_1(x_2, \ldots, x_n)x_1^{d-1}+\ldots +a_d(x_2,\ldots , x_n)$ with $\deg a_i\leq i$. This works for every infinite base field, see \cite[Lemma 2.1.1]{Kempf}. $\blacksquare$
\section{Generic exactness of 3D SUSY}
\label{exatagen}

The following result is general and holds for a number $\mathrm{N}\geq 2$ of superfields, the case $N=1$ being trivial.
\begin{theorem}[Main Result on Exactness]\label{susyexactN} Given $N\geq 2$, outside a real hypersurface $H$ in $E_N$, every superpotential $\Phi\in E_N$ gives rise to an exact (simple) SUSY in 3D.  Equivalently, if the homogenised components of the gradient of $\Phi$ have no common zeros at infinity, then $\Phi$ has a real critical point.  In particular, 3D simple SUSY with N real superfields is generically exact.  \end{theorem}
{\bf Proof:} We shall assume that $\Phi\in E_N$ is a proper quartic superpotential in $N$ variables (i.e. of proper degree $4$, so $\Phi$ lies outside its subspace of cubic functions).  Let $\tilde{\Phi}$ be its homogenization, $X_0=0$ being $\pp_{\infty}^N$.

  Let $F_i=\partial_{X_i}\tilde{\Phi}$, $G_i=F_i(0, X_1, \ldots, X_N), 1\leq i\leq N.$ By Proposition \ref{discr}, outside a real hypersurface in $E_N$ we may assume that $G_i$ have no common zeros, which in turn means by Proposition \ref{liouville} that the zero set of $F_1, \cdots, F_N$ is finite.   By B\'ezout's Theorem, the total sum of multiplicities equals
	$$\sum n_{P_i}=3^N.$$
  By Lemma \ref{realmult}, a complex pair of common zeros $P\neq\overline{P}$ contributes an even number to the above sum, which necessarily means that there is a real zero $Q$ of odd multiplicity.  Such $Q$ is a real point in $X_0\neq 0$ (i.e. $Q$ lies in the affine part), and corresponds to a critical point of $\Phi$. Those $\Phi\in E_N$ satisfying this are indeed generic, by Proposition \ref{zeromeasure}. $\blacksquare$

\section{Explicit family with F-term breaking}
\label{explicitbreaking}

\begin{theorem}\label{familiapasso1} Let $a, D\neq 0$ be real parameters, and let $A(u,v)$ be a quartic homogeneous polynomial.  Define the superpotential  $\Phi=\Phi_1+\Phi_4$, where 
	$$\Phi_4(x,y,z)=A(x+ay,z)$$
	and   $\Phi_1(x,y,z)=Dx.$ The gradient $\nabla\Phi=(F,G,H)$ is non-vanishing in $\R^3$ (i.e. $\Phi$ has no critical points).  The above described superpotentials forms a $6$-parameter family exhibiting spontaneous breaking of SUSY ($N=3$ superfields).
\end{theorem}
{\bf Proof:}  Consider the equations $\partial_x\Phi=\partial_y\Phi=0,$ which in their explicit form are:
$$\partial_uA=0, \quad D+a\partial_uA=0,$$
which is clearly impossible for $D\neq 0$.  This produces a $6$-parameter example, counting $D$ and the coefficients of $A$.  The Theorem is thus settled. $\blacksquare$

\begin{caja}{Remark:} The family obtained in Theorem \ref{familiapasso1} is indeed not very big. Non-existence of affine real solutions relies on the fact that every complex common zero $P_i$ of the homogenised versions of $F,G,H$ lies at infinity, and the (projective) intersection cycle equals $3\sum P_i$ ($\sum P_i$ being the intersection cycle of the plane curves $\tilde{F}(0,X_1,X_2,X_3),\tilde{H}(0,X_1,X_2,X_3)$) if the intersection be proper.  Indeed, the monomial $Z_0^3$ is a linear combination of $\tilde{F}, \tilde{G}$. The rest follows from Section \ref{intro}.\end{caja}

\section{The case of `almost-even' superpotentials}
\label{evenpot}
Let $\Phi$ be even.  This means that the cubic polynomials $F,G,H$ are odd, and of the form
$$F=F_1+F_3, G=G_1+G_3, H=H_1+H_3.$$
We wish to study the cases where no solution other than the origin exists.

  An even superpotential necessarily has the origin as a critical point, so it is not the answer we seek.  However, it is the starting point for our approach.

\begin{theorem}\label{quebrayuri}There exists a $18$-dimensional family of quartic superpotentials on $\R^3$, $\Phi\in E_3$, with no critical points in $\R^3$.\end{theorem}
{\bf Proof:}

 Assume that there is an even quartic superpotential $V\in E$, with homogeneous pieces $V=V_2+V_4$, where $V_2$ is positive definite and $\partial_{X_2}V_4=X_1 Q(X_1,X_2,X_3)$, $Q$ positive definite.  Let $\tilde{V}=X_0^2V_2(X_1,X_2,X_3)+V_4(X_1,X_2,X_3)$ be its homogenisation.  The derivatives are:
$$K=\partial_{X_0}\tilde{V}=2X_0V(X_1,X_2,X_3),$$
$$\tilde{F}=X_0^2\partial_{X_1}V_2+\partial_{X_1}V_4,$$
$$\tilde{G}=X_0^2\partial_{X_2}V_2+X_1Q(X_2,X_3)$$
and $\tilde{H}=X_0^2\partial_{X_3}V_2+\partial_{X_3}V_4$.

Consider the equations $K=0, \tilde{G}=0, \tilde{H}=0$.  The intersection  $X_0=0, \tilde{G}=0, \tilde{H}=0$ corresponds to the intersection of two cubic plane curves on $X_0=0$, one of which is $X_1Q(X_2,X_3)$, and the only real common zeros of $K, \tilde{G}, \tilde{H}$  lie on the plane $X_1=0$.

 Take now the function $\Phi(u,v,w)=\tilde{V}(u,1,v,w)$.  The function $\Phi$ has no critical points, since the dehomogenisations of $K, \tilde{G}, \tilde{H}$ are the partial derivatives of $\Phi$. The two positive definite quadrics, the constant multiples of $X_1^aX_3^{4-a}$ ($0\leq a\leq 4$) appearing after integrating $\partial_{X_2}\tilde{V}=X_1Q(X_1,X_2,X_3)$ w.r.t. $X_2$ and the value $\Phi(0,0,0)$ total up to $18$.  $\blacksquare$

\begin{caja}{Remark:} The number of parameters of this family is still much lower than 34. One may see that we have excess intersection in this family, i.e. the intersection is not proper over $\C$, which prevents us from enlarging it into a bigger family.  See Section \ref{SUSYquebra32} for a successful attempt.\end{caja}

\subsection{Final Ansatz}

\label{finalansatz}

\begin{theorem}\label{N2} ($E_N, N=2$) There is a quartic polynomial $\Phi(u,v)\in E_2$ with no critical points, and such that the projectivisations $\tilde{F}=0,\tilde{G}=0$ of the real curves $\partial_u\Phi=0, \partial_v\Phi=0$ have a unique real point of intersection, with multiplicity, one, which lies at infinity. \end{theorem}
{\bf Proof:} Let $U(X_0, X_1, X_2)=X_0^2V_2(X_1,X_2)+V_4(X_0, X_1, X_2)$, where $V_2$ is a positive definite binary quadratic form and $V_4$ is such that $\partial_2U=\lambda X_1(X_1^2+bX_1X_2+cX_2^2)$, where $b^2-4c<0, \lambda\neq 0$, where $V_2$ and $X_1^2+bX_1X_2+cX_2^2$ are assumed to have no common factors. Its partial derivatives $\partial_0U, \partial_2U$ are
$$\partial_0U=2X_0 V_2(X_1,X_2), \partial_2U=X_1(X_1^2+bX_1X_2+cX_2^2).$$
 By \cite[Sec. 3.3]{Fulton} The intersection of both curves is an effective cycle of degree $9$, with only a real point, which has multiplicity $1$, i.e. $(0:0:1)$ for $X_0=0, X_1=0$.  Thus, the superpotential $\Phi(u,v)=U(u,1,v)\in E_2$ has no critical points, as desired. $\blacksquare$

\begin{cor} \label{Narbitr} Given $N\geq 3$, there exists a quartic superpotential $\Psi\in E_N$ without critical points (of homogenization $\tilde{\Phi}$) such that the intersection cycle $F_1\bullet \cdots \bullet F_N$ where $F_i=\partial_i\tilde{\Phi}$ (partial derivatives of its homogenization) has a real point of multiplicity one at infinity, and no other real points in $\pp^N_{\R}$.\end{cor}
{\bf Proof:} Indeed, consider the example $\Phi$ provided in Theorem \ref{N2} and define 
$$\Psi(u_1, \ldots, u_N)=\Phi(u_1,u_2)+\sum_{i=3}^N (u_i^2+u_i^4).$$
The intersection cycle is as desired. $\blacksquare$

\section{A sharp result on SUSY breaking}
\label{SUSYquebra32}

  The existence result in Corollary \ref{Narbitr} brings about a $34-$parameter family out of the $35$ parameters defining the superpotential $\Phi$ (in the case of 3 superfields).  In fact, family of  maximal dimension showing F-term SUSY breaking may be produced for 3 superfields, and likewise for $N\geq 3$ superfields, over a small ball in $\R^{34}$, as we shall prove.  Since the quartic superpotentials with exact SUSY form the complement of a real hypersurface and of the subspace of cubic polynomials, this result is optimal. We cannot describe, however, what happens outside such small ball in $H$, i.e., we could not find a nice description for the precise locus of superpotentials within $H$ where spontaneous breaking holds at their F-term.

  The following lemma holds more generally, but we need it only in the case of finite zero sets.
\begin{lemma} \label{proper} Let $U^{(0)}(X_0, \ldots, X_n)$ be a complex form of degree $d>0$, and assume that the system of equations $\partial_1U^{(0)}=0, \ldots, \partial_nU^{(0)}=0$ has a finite number of solutions.  Fix the geodesic distance on each $\pp^n_{\C}$ with respect to the Fubini-Study metric.  The distance between $\mathcal{Z}^{(0)}$ and the zero set of a degree-$d$ form  $U$ sufficiently close to $U^{(0)}$ is small. \end{lemma}
{\bf Proof:} Let $\nu_d=\binom{n+d}{d}-1$, and let $\pp^{\nu_d}$ denote the projectivisation of the vector space $\C[X_0, \cdots, X_n]_d$, to which $U^{(0)}$ belongs; we denote by $[U]$ the class of $U$ in its projectivisation. Consider the distance function  on the product $\pp^r_{\C}\times \pp^s_{\C}$ to be the sum of geodesic distances on every factor (in fact, we need only a distance that is compatible with the topology given).  The composition
$$\pi:\{(p,[U]):\partial_1U(p)=0, \ldots, \partial_nU(p)=0\}\subset \pp^n\times \pp^{\nu_d}\twoheadrightarrow \pp^{\nu_d}$$
is a proper map, hence closed, and so for every $\epsilon>0$, $\pi^{-1}(B(U^{(0)},\epsilon)$ contains an open set of the form $\{d(Z^{(0)},(p,U))<\eta \}$ for some $\eta=\eta(\epsilon)>0$. $\blacksquare$

\begin{theorem}\label{quebramax} Let $N=3$.  Consider the superpotential $\Phi^0$ obtained in the proof of  Corollary \ref{Narbitr}, and let $\tilde{\Phi}^0$ be its homogenization.  There is a small neighbourhood in $E$ around $\Phi^0$ and a real-analytic hypersurface therein which provides a family of superpotentials with SUSY breaking at the F-term. Such family has 34 parameters, and represents a small neighbourhood within the algebraic hypersurface from Theorem \ref{susyexactN}.

  By analogous arguments, one obtains a codimension-1 family in $E_2$ around the superpotential $\Phi^0$ in Theorem \ref{N2} with F-term supersymmetry breaking with $N=2$ superfields.
\end{theorem}
{\bf Proof:} 
 Consider the following four equations in $p\in\pp^3_{\R}$ and  $[\Phi]\in\pp(E)$:
$${\mathcal M}=\{(p,[\Phi]):\partial_1\tilde{\Phi}(p)=0,\, \partial_2\tilde{\Phi}(p)=0,\, \partial_3\tilde{\Phi}(p)=0, \, p\in H_\infty,$$

and by the Implicit Function Theorem, the equations $\partial_i\tilde{\Phi}(p)=0$ ($i=1,2,3$) define a locus which projects diffeomorphically around $(p^0,\tilde{\Phi}^0)$ onto an open neighbourhood $B_0$ of $[\Phi^0]$ in $\pp(E)$.  By construction, the corresponding polynomials $\Phi$ in such pairs $(p,\tilde{\Phi})$ are already in the hypersurface $H$ from Theorem \ref{susyexactN}.
% puedes poner aquí eso ...
  By Lemma \ref{proper} and analogous arguments, if the neighbourhood is sufficiently small then the other zeros of $\partial_i\Phi=0, 1\leq i\leq 3$ remain complex (consider the distance function to $\pp^n_{\R}$ as well).  Let us get back to the equation $p\in H_{\infty}$: this produces a real-analytic hypersurface $\cH \subset B_0\subset \pp(E)$, and the cone over $\cH$ (i.e. $p^{-1}(\cH)$, where $p:E-\{0\}\to \pp(E)$ is the canonical projection) provides the required family in $E$ itself.  The rest is clear.  $\blacksquare$

  An entirely analogous argument yields an N-superfield analogue.
\begin{theorem}\label{quebramaxN} Given $N\geq 2$, there is a quartic superpotential in $N$ variables $\Phi^1$ with no critical points in $\R^N$, which gives rise to a small ball of dimension $\dim E_N-1$ around $\Phi^1$ in $E_N$ satisfying this condition (spontaneous F-term breaking of 3D SUSY with maximal dimension). 
\end{theorem}
{\bf Proof:} Take $\Phi^0$ as in Corollary \ref{Narbitr}, and proceed as in the proof of Theorem \ref{quebramax}. $\blacksquare$

\section{Concluding Comments}
\label{conclusion}
To conclude, we have presented a general result on the exactness of simple SUSY in (1+2)-D: F-type spontaneous breaking is not realisable. This is basically due to the real structure underneath non-extended SUSY in 3D, contrary to 4D-SUSY, for which the underlying structure in complex. Also, in 3D-SUSY another key feature is that renormalisability allows scalars to have sixth power self-interactions, contrary to 4D, where scalar self-interaction are only quartic. This drastically changes the scenario in going from 4D into 3D. We have found a family of 6-th order scalar superpotentials with maximal number of free parameters, featuring F-term spontaneous breaking of 3D SUSY in the case of $N\geq 2$ superfields. However, only small balls around quite explicit superpotentials have been obtained, without showing the global shape of the F-term breaking locus.  An explicit, simpler albeit `smaller' family in the case of 3 superfields has been provided in Theorem \ref{familiapasso1}.  In the latter case, the real zeros lie at infinity, but are multiple, which prevents from perturbing to bigger families --we managed to do so by using a point of multiplicity one in Theorem \ref{quebramaxN}.
  In our case, the locus of F-term spontaneous breaking within the parameter space of space-time configurations of $N\geq 2$ superfields is contained in a real (algebraic) hypersurface $H$, and contains at least a small neighbourhood in $H$, whereas in 4D SUSY \cite{ORafer} its corresponding F-term breaking locus (only for $N\geq 3$) is contained within a complex hypersurface (real codimension two), as our proof of Theorem \ref{susyexactN} and the reference therein show.  This phenomenon is explained by the fact that simple SUSY in 4D corresponds to double SUSY in 3D. Contrary to the 4-dimensional case, where a non-renormalization theorem holds for the superpotential, simple SUSY is not protected in 3D against quantum corrections. Thus, if SUSY is not broken at the tree-level, quantum corrections might, in general, yield SUSY breaking. In 4D, however, SUSY breaking is exclusively a tree-level effect.

% \bibitem{c}
% Author, \emph{Title},
% Publisher (year).

% Please avoid comments such as "For a review'', "For some examples",
% "and references therein" or move them in the text. In general,
% please leave only references in the bibliography and move all
% accessory text in footnotes.

% Also, please have only one work for each \bibitem.

\end{document}